\documentclass{INTERSPEECH2023}
\usepackage{booktabs, multirow}
\usepackage[para]{footmisc}

\interspeechcameraready

\title{ControlVC: Zero-Shot Voice Conversion with Time-Varying Controls on Pitch and Speed}
\name{Meiying Chen and Zhiyao Duan}
\address{
  Department of Electrical and Computer Engineering, University of Rochester, Rochester, NY, USA
\email{meiying.chen@rochester.edu, zhiyao.duan@rochester.com}
}

\begin{document}

\maketitle


\begin{abstract}
Recent advancements in neural speech synthesis have renewed interest in voice conversion (VC) to go beyond timbre transfer. Achieving controllability of para-linguistic parameters like pitch and speed is crucial in various applications. However, existing studies either lack interpretability or only provide global control at the utterance level. This paper introduces ControlVC, the first neural voice conversion system to enable time-varying controls on pitch and speed. ControlVC uses pre-trained encoders to generate pitch and linguistic embeddings, combined and converted to speech using a vocoder. Speed control is achieved by TD-PSOLA pre-processing, while pitch control is achieved by manipulating the pitch contour before feeding it into the encoder. Systematic subjective and objective evaluations show that this work significantly outperforms self-constructed baselines on speech quality and controllability for non-parallel zero-shot conversion while achieving time-varying control \footnote{
Proceedings of Interspeech 2023, doi: 10.21437/Interspeech.2023-1788
\newline This work is partially supported by New York State Center of Excellence in Data Science award and synergistic activities funded by the National Science Foundation (NSF) under grant DGE-1922591. 
A special thanks to Yongyi Zang for developing the subjective evaluation website. 
\newline Code and audio samples: \url{https://bit.ly/3PsrKLJ}}.

\end{abstract}

\section{Introduction}
\label{sec:intro}
Voice conversion (VC) is the task of alternating the timbre, style and other para-linguistic features of a speech utterance while maintaining its linguistic content \cite{Review}. It is gaining increasing attention from researchers in various domains, thanks to its broad applications in human-computer interaction, virtual human, and multimedia production. Existing VC systems focus on the conversion of timbre and style of the source speaker to those of a target speaker \cite{Review, vcc20}. The controllability of other para-linguistic features such as pitch and speed, however, has not received much attention. In speech communication, para-linguistic features are critical in conveying the emotion, intention and even semantic meaning of the talker \cite{emotionbook}. 
For example, raising the pitch and slowing down help to emphasize a word \cite{Carlson1989}.
Therefore, achieving controllability on para-linguistic features such as pitch and speed is a critical step toward making VC techniques useful in many application scenarios.

In general, there are two levels of control on para-linguistic features in VC. \emph{Global control} refers to controls at the utterance-level, and is often realized under the umbrella of \emph{style transfer}. For example, when an utterance is converted from a male speaker to a female speaker, the overall pitch range is often raised. 
Global control has been well achieved in many modern VC systems \cite{qian2020f0, RN5, li2021starganv2}
\emph{Local control} refers to time-varying controls of para-linguistic features. 
Little attention has been paid to it by modern neural-based methods, and it is the concern of this paper.

There are two major categories of VC systems: \emph{parametric methods} and \emph{end-to-end methods}. Parametric methods first apply a statistical model or a neural network to estimate speech parameters from the source and target utterances, and then use these parameters to generate converted speech with a vocoder~\cite{chen2010gmm, sun2015voice}.
These methods generally can provide good controllability by modifying the explicitly predicted parameters, but limited performance on transferring the target timbre due to insufficient capacity~\cite{nercessian2021end}. 

In recent years, end-to-end VC methods have shown significantly better performance on timbre transfer and speech naturalness of the converted utterance \cite{kameoka2018stargan, li2021starganv2, lin2021fragmentvc, casanova2022yourtts, nguyen2022nvc}.
However, the controllability on para-linguistic features is sacrificed as they are stored in network weights that are difficult to interpret. 
More recent works try to disentangle different aspects of speech such as content, timbre, pitch and speed into separate embeddings to achieve controllability, however, such embeddings, and the controls of them, are often at the global level \cite{RN14, RN8}.
Even with frame-level embeddings like those in \cite{RN18}, time-varying control is still difficult to realize, since the embeddings do not have an explicit mapping to human-interpretable parameters of pitch and speed, and the influences of such embeddings on the generation is not clear.

A natural thought is to design a cascade system to achieve time-varying (local) control on pitch and speed in voice conversion: First apply a neural-based method for timbre transfer, and then apply signal processing methods such as time-domain pitch synchronous overlap and add (TD-PSOLA) \cite{RN11} to perform time-stretching and pitch shifting.
We tried this, but observed significant artifacts in the converted utterance. We argue that this is because the pitch and speed controls can be hardly designed natural, and there is no following steps in the VC pipeline to fix this unnaturalness. 
For example, when one speeds up, consonants and vowels are sped up at different rates depending on the context. Similarly, when one raises the pitch, different phonemes are raised at different degrees.

In this paper, we propose ControlVC, a voice conversion system that achieves time-varying control on pitch and speed. 
ControlVC performs speed control by modifying the speed of the source utterance using TD-PSOLA. It then performs pitch control by modifying the pitch contour of the speed-controlled source utterance, and uses a VQ-VAE pitch encoder to compute discrete pitch embedding. A pre-trained HuBERT extracts the linguistic embedding from the speed-controlled source utterance, and a pre-trained speaker encoder extracts the speaker embedding from the target utterance.
A modified version of the HiFi-GAN vocoder \cite{hifigan} is then used to generate the waveform of the converted utterance by integrating the pitch, linguistic and speaker embeddings. It is noted that the pre-trained speaker encoder enables the model to generalize to unseen speakers, and the pre-trained linguistic encoder using vast datasets improves the coverage of diverse linguistic content. They work together to help the system to work in a zero-shot conversion scenario, without the need of additional training data from the source or target speaker.

To our best knowledge, ControlVC is the first neural VC system that achieves time-varying controls on pitch and speed. 
ControlVC performs a generic control on pitch and speed but not trying to mimic those of the target speaker.
It's worth noting that works such as \cite{lee2022duration, lian2021towards, choi2021neural} may have the potential to achieve time-varying speed and/or pitch control, but substantial modifications of the algorithms and necessary experiments would be needed to support such claim, which is not the primary focus of their research. 
To validate the proposed system, we conduct extensive subjective and objective evaluations and compare it with two self-constructed baselines as no existing systems were available. 
Experimental results show that ControlVC realizes a good level of time-varying controllability on pitch and speed, while achieving significantly better naturalness and timbre similarity than the comparison methods.

\section{Proposed ControlVC System}

\label{sec:proposed}
\label{sec:proposed: overview}
\begin{figure}
  \centering
  \includegraphics[width=1\columnwidth]{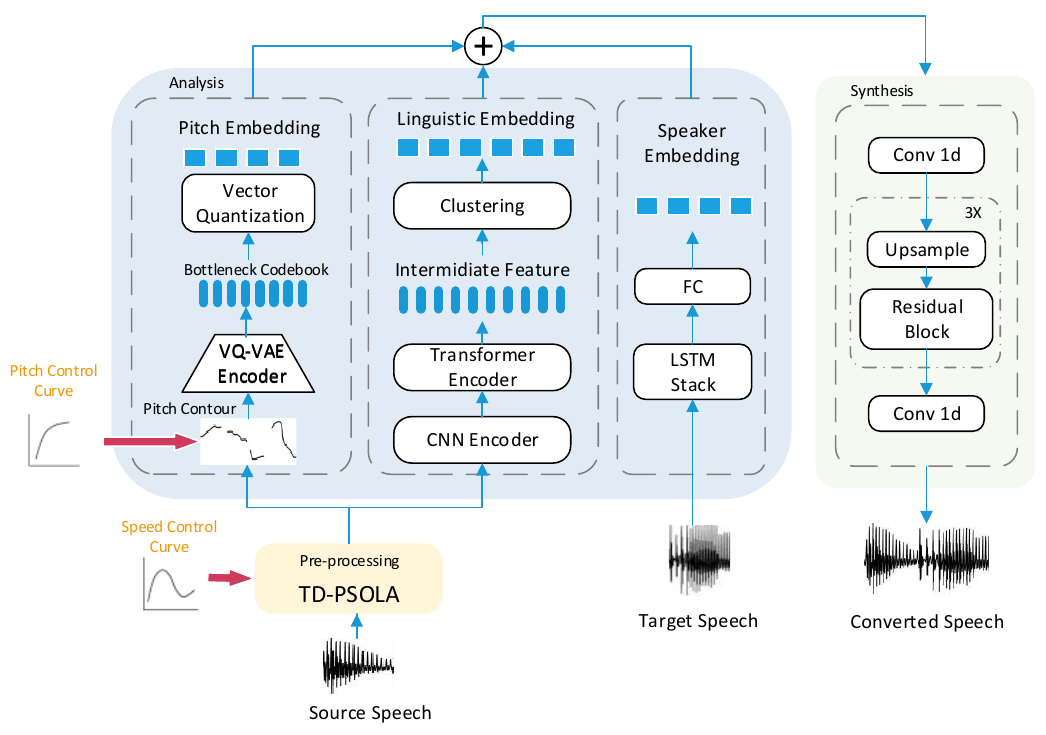}
  \caption{System Overview.}
  \label{fig:overview}
\end{figure}

\subsection{Overview}

ControlVC aims to achieve time-varying control over pitch and speed in non-parallel and zero-shot voice conversion using control curves. As shown in Figure \ref{fig:overview},
The system consists of three stages: pre-processing, analysis and synthesis. 
The pre-processing stage employs TD-PSOLA to modify the speed of the source speech according to the speed control curve. 
In the analysis stage, the pitch contour of the processed source utterance is estimated and modified by the pitch control curve, before being fed into a VQ-VAE to obtain a pitch embedding. A linguistic embedding is computed from the speed-modified source utterance through a linguistic embedding network. Finally, a speaker embedding is computed from the target utterance. 
The pitch, linguistic and speaker embeddings are then up-sampled, concatenated and fed to the synthesis stage, which uses HiFi-GAN neural vocoder~\cite{hifigan} to synthesize the time-domain waveform of the converted speech utterance. 
Note that only the HiFi-GAN vocoder is trained from scratch on the voice conversion task, while the linguistic, speaker, and pitch encoders are pre-trained on other tasks and fixed.


\subsection{TD-PSOLA Prepossessing and Speed Control}

In the preprocessing stage, we use the time-domain pitch synchronous overlap and add (TD-PSOLA) algorithm \cite{RN11} to modify the speed of the source utterance according to the input speed control curve. 
We first segment the original utterance and apply time-stretching to each frame using the stretching ratio indicated by the control curve at the corresponding location. The pitch is retained, and so are the timbre and linguistic content. 

\subsection{Pitch Control and Pitch Embedding}

We employ the YAAPT algorithm \cite{kasi2002yet} to extract the pitch sequence $(p_{1} ,\cdots, p_T)$ of the speed-controlled source utterance with a frame length of 20 ms and a hop size of 5 ms, where $T$ is the number of frames. This pitch sequence is then multiplied by the input pitch control curve to obtain the modified pitch sequence $(p'_{1} ,\cdots, p'_T)$.
The modified pitch sequence is fed to a VQ-VAE based pitch embedding network \cite{dhariwal2020jukebox} to obtain the pitch embedding for the converted utterance.
Taking the pitch sequence as input, the encoder produces a sequence of 128-d latent vectors $(\mathbf{h}_{1}, \cdots, \mathbf{h}_T)$, which are then mapped to their respectively closet codes in a bottleneck codebook.
We then take the integer indices of the codebook vectors to form the pitch embedding sequence$\mathbf{z}^{(p)}=(z^{(p)}_1, \cdots, z^{(p)}_T)$.

This VQ-VAE embedding network is trained on original utterances in the training set without applying speed and pitch controls, by minimizing the mean squared error (MSE) between the estimated pitch sequence and the original pitch sequence.

\subsection{Linguistic Embedding}

To maintain the linguistic content of the source utterance, we need a linguistic encoder to compute the linguistic embedding from the (speed-controlled) source utterance. 
We use a publicly available HuBERT model that is pre-trained on 960 hours of LibriSpeech audio. In our system, the input to the linguistic encoder is the source waveform, segmented into the same frames as those fed to the pitch detector. 
The outputs are 768-d feature vectors extracted from the 6-th layer, one vector for each frame. 
As the feature vectors extracted from HuBERT are continuous and may contain speaker information, a K-means clustering procedure is applied on the HuBERT output. We train a mini-batch K-means clustering algorithm with scikit-learn 
on the LibriSpeech-train-clean-100 dataset. During voice conversion model training, new data is assigned to pre-stored clusters based on the distance to the centroids.
The final linguistic embedding is the sequence of the integer cluster indices of each frame $\mathbf{z}^{(l)}=(z^{(l)}_1, \cdots, z^{(l)}_T)$, where $K$ is set to 100.

\subsection{Speaker Embedding}
In order to transfer the timbre information of the target speaker, we use a speaker encoder to compute the speaker embedding from the target utterance. 
Following the design of \cite{RN12}, our speaker encoder has a stack of two LSTM layers with 768 cells. It takes Mel-spectrogram as input and passes the outputs of the last time step through a fully connected layer. This results in a 256-d speaker embedding vector, which is then copied into a embedding sequence $\mathbf{z}^{(s)}$, to match the same frame rate as that of the pitch and linguistic embeddings. 

\subsection{HiFi-GAN Neural Vocoder}
To construct the encoded discrete representation, the linguistic and pitch embeddings are up-sampled, and the utterance-level speaker embedding is copied to match the same frame rate. These three embeddings are then concatenated into the intermediate representation $\mathbf{z} = (\mathbf{z}^{(p)}, \mathbf{z}^{(l)}, \mathbf{z}^{(s)})$, which is then fed into a neural vocoder to generate the waveform.
We use HiFi-GAN's \cite{hifigan} vocoder but modify the original implementation so that it directly accepts the discrete and continuous mixed representation $\mathbf{z}$ as input. 
The generator includes a set of transposed convolution blocks to increase sample rate of the discrete representation and a residual block with dilated layers to increase the receptive filed. The discriminator contains two types of sub-discriminators:  five multi-period discriminators (MPD) and three multi-scale discriminators (MSD). 

We denote the generator as $G$ and the discriminator as $D$, which contains a total of $K=8$ sub-discriminators as $D_k$, for $k \in {1, \cdots, K}$. 
The objectives for training the generator and the discriminator are:
\begin{align}
    L_{G} &= \sum_{k=1}^K \biggl{[}L_{Adv}(G; D_{k}) +  \lambda_{fm}L_{FM}(G; D_{k}) \biggr{]} \nonumber \\ &+ \lambda_{mel}L_{Mel}(G),\\
    L_{D} &= \sum_{k=1}^K  L_{Adv}(D_{k}; G),
\end{align}
where $L_{Adv}$, $L_{FM}$ and $L_{Mel}$ are adversarial loss, feature matching loss and mel-spectrogram loss, respectively.
Following \cite{hifigan}, the tradeoff parameters $\lambda_{fm}$ and $\lambda_{mel}$ are set to 2 and 45, respectively. The feature matching loss $L_{FM}$ and mel-spectrogram loss $L_{Mel}$ are defined as:
\begin{align}
    L_{Mel}(G) &= \mathbb{E}_{x, \hat{x}} \biggl{[} ||\phi(x) - \phi(G(z(x)))||_{1} \biggr{]},
\end{align}
where $x$ is the ground-truth audio. $M$ denotes the number of layers in the discriminator. $D^{i}$ and $N^{i}$ represent the features and the number of features in the $i$-th layer, respectively.
The function $\phi$ transforms the waveform to a mel-spectrogram.

\section{Experiments}
\subsection{Dataset}

We evaluate ControlVC on the CSTR VCTK Corpus \cite{RN19}, which includes 44 hours of clean speech uttered by 110 English speakers with various accents. 
All recordings are downsampled to 16 kHz.
We randomly select 10 speakers (5 male and 5 female) and use all of their utterances for testing, and the remaining 100 speakers for training. 
In total, there are 39,781 utterances in the training set and 3,690 utterances in the test set.

\subsection{Baseline Methods}

To our best knowledge, ControlVC is the first VC method that achieves time-varying control on pitch and speed, and no existing methods were found for direct comparison. As noted earlier, while \cite{lee2022duration, lian2021towards, choi2021neural} may have the potential to achieve time-varying control, substantial modifications of their methods are needed.
Therefore, we designed two baselines using well-established algorithms in signal processing and VC to achieve time-varying control.
Note that both baselines are new controllable VC systems that do not exist in the literature.

The first baseline, named \emph{P-LPC}, employs TD-PSOLA to modify the pitch and speed of the source utterance, and linear predictive coding (LPC) \cite{zolzer2002dafx} to model the timbre of the target speech and transfer it to the converted utterance.
The second baseline, named \emph{P-AutoVC}, first uses TD-PSOLA to modify the pitch and speed of the source, and then uses AutoVC \cite{RN12} to achieve timbre transfer. AutoVC is a widely-used neural-based VC method that achieves high audio quality but no pitch or speed controllability. 

\subsection{Training}
For the proposed ControlVC method, the pitch encoder is pre-trained on the VCTK dataset for 40k steps. 
The linguistic embedding is extracted from the 6-th layer of a publicly available pre-trained HuBERT model \cite{fairseq}.
The speaker encoder is pre-trained on a combination of VoxCeleb \cite{Nagrani17} and Librispeech \cite{RN2} datasets with a total of 3,549 speakers using GE2E loss \cite{autovccode}.
Finally, we train the HiFi-GAN vocoder on the VCTK dataset using one RTX 2080Ti with batch size 8 for 350k steps. We use Adam optimizer with an initial learning rate of 0.0002 and a decay rate of 0.999. For the PSOLA-AutoVC baseline, we use the pre-trained AutoVC model available online \cite{autovccode}.

\subsection{Experimental Setup}
The VC experiments are performed among all 90 pairs of 10 test speakers. Each utterance of one test speaker is converted to each of the other 9 speakers' voices. 
Each test speaker reads a different set of sentences. All the 10 speakers and their sentences are unseen during training.
Therefore, the conversion is non-parallel and zero-shot.

In this experiment, we apply control curves for speed and/or pitch. 
Four control settings are tested: ``No Control" - traditional voice conversion without any explicit control; ``Pitch Only" and ``Speed Only" denote pitch or speed control but not both; ``Speed+Pitch" means both aspects are controlled.
We test two curves for pitch control: \emph{stressing} (i.e., pitch rising abruptly then going down gradually) and \emph{rising}, and three curves for speed control: \emph{parabola}, \emph{speed up} and \emph{slow down}.
The control settings and the control curves are drawn with equal probability for each conversion.
We perform both subjective and objective evaluations to assess the conversion quality, intelligibility and controllability of the proposed system. 

\subsection{Objective Evaluation}
We conduct an objective evaluation to assess the speech intelligibility, timbre similarity and controllability of the converted utterances.
For speech intelligibility evaluation, we use IBM speech recognition service \cite{RN15} to transcribe converted speech into text and then calculate the word error rate (WER) \cite{wang2003word} against the ground-truth transcripts.
For timbre similarity, we first use a pre-trained speaker encoder Resemblyzer \cite{Resemblyzer} to extract speaker embeddings of the converted and the target utterance. Then we score the speaker similarity (Sim.) by calculating the cosine distance between the embeddings on a scale of 0 to 1, the higher the more similar.
The same set of samples generated for subjective evaluation is used for this section.

Table~\ref{obj} shows the objective evaluation results. It can be seen that ControlVC outperforms both baselines by a large margin on both metrics across all test configurations, including no control, pitch or speed only control and pitch+speed control, showing its superior performance on audio quality and intelligibility in various controllable conversion settings. 

\begin{table}[]
\centering
\caption{Objective evaluation results.}
\begin{tabular}{cc|c|c}
\hline
                                                          &                   & \textbf{Sim. ↑} & \textbf{WER (\%) ↓} \\ \hline
\multicolumn{2}{c|}{\textbf{GT}}                                              & \textbf{1.00}   & \textbf{9.82}       \\ \hline
\multicolumn{1}{c|}{\multirow{3}{*}{\textbf{No Control}}} & P-LPC    & 0.65            & 89.12               \\ \cline{2-4} 
\multicolumn{1}{c|}{}                                     & P-AutoVC & 0.66            & 76.51               \\ \cline{2-4} 
\multicolumn{1}{c|}{}                                     & Proposed & \textbf{0.85}   & \textbf{10.99}      \\ \hline
\multicolumn{1}{c|}{\multirow{3}{*}{\textbf{Pitch Only}}} & P-LPC    & 0.66            & 88.56               \\ \cline{2-4} 
\multicolumn{1}{c|}{}                                     & P-AutoVC & 0.65            & 72.83               \\ \cline{2-4} 
\multicolumn{1}{c|}{}                                     & Proposed & \textbf{0.82}   & \textbf{12.40}      \\ \hline
\multicolumn{1}{c|}{\multirow{3}{*}{\textbf{Speed Only}}} & P-LPC    & 0.65            & 88.64               \\ \cline{2-4} 
\multicolumn{1}{c|}{}                                     & P-AutoVC & 0.65            & 72.59               \\ \cline{2-4} 
\multicolumn{1}{c|}{}                                     & Proposed & \textbf{0.84}   & \textbf{16.37}      \\ \hline
\multicolumn{1}{c|}{\textbf{Pitch+Speed}}                 & Proposed & \textbf{0.83}   & \textbf{22.46}      \\ \hline
\end{tabular}
\label{obj}
\end{table}
\subsection{Subjective Evaluation}

We perform two subjective experiments using a self-designed survey website which is publicly available and shared within the University of Rochester and its alumni to recruit study participants without providing monetary incentives. 

\begin{figure}[t]
  \centering
  \includegraphics[width=0.9\columnwidth]{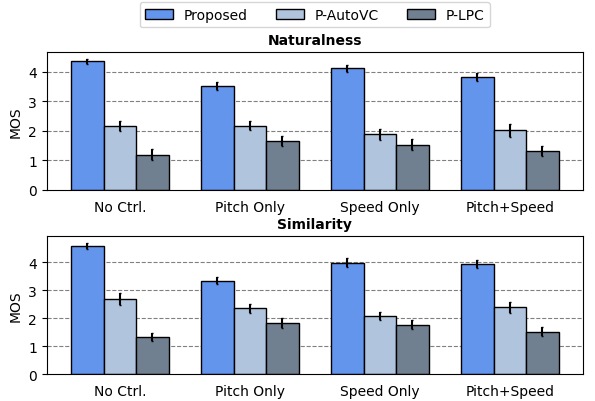}
\caption{MOS results on audio quality (naturalness and timbre similarity) with 95\% confidence intervals.} 
\label{fig:tabl1}
\end{figure}

\textbf{Audio Quality Test.} In the first test, we use mean opinion score (MOS) \cite{streijl2016mean} to assess the naturalness and timbre similarity of the converted speech. 
Study participants are presented with a set of utterances including one source, one target and several converted utterances. Each set is referred to as a sample. 
The converted utterances from ControlVC and the two baselines are presented in random order.
For each sample, participants are asked to rate between 1-5 on the naturalness of the converted utterances and the timbre similarity between the target and the converted utterance.  
Higher scores are better.
Each participant is asked to complete at least 6 samples, i.e., 36 ratings for 18 converted utterances from 3 comparison methods. Participants are allowed to complete more samples.
In total, 233 samples are evaluated, resulting in 1398 ratings.
As shown in Fig. \ref{fig:tabl1}, ControlVC archives the best MOS among three comparison methods in all control settings. In addition, comparing the three with-control settings with ``No Control'', we see that applying controls only slightly decreases the speech quality of the converted speech from ControlVC.

Note that for unseen-to-unseen zero-shot voice conversion, the original AutoVC paper~\cite{RN12} reports MOS scores on naturalness and similarity of about 3.1 and 2.9, and ~\cite{choi2021neural} reports MOS on naturalness of 2.59 using AutoVC model.
However, our P-AutoVC baseline only achieves 2.17 and 2.68 MOS on naturalness and similarity. In addition to consistent subject biases, we suggest that there may be other reasons. First, in ~\cite{choi2021neural} AutoVC model is trained on VCTK (44 hours) and LibriTTS (360 hours), while in our work P-AutoVC is trained on VCTK only to ensure a consistent training setup with the proposed system.
In addition, AutoVC is applied after the PSOLA preprocessing, which introduces noticeable artifacts that are very likely to affect the performance. Due to both reasons, we believe that the performance degradation of P-AutoVC baseline is reasonable, which also explains its subpar performance in Table \ref{obj}.

\begin{table}[]
\caption{MOS results on controllability with 95\% confidence intervals.}
\centering

\begin{tabular}{cccc}
\hline
\multicolumn{4}{c}{\textbf{Controllability Rating}}                                                                                                                                                                \\ \hline
\multicolumn{2}{c|}{}                                                                                                                           & \multicolumn{1}{c|}{\textbf{Pitch}}       & \textbf{Speed}       \\ \hline
\multicolumn{1}{c|}{\multirow{2}{*}{\textbf{\begin{tabular}[c]{@{}c@{}}Pitch \\ Only\end{tabular}}}} & \multicolumn{1}{c|}{Real Curve} & \multicolumn{1}{c|}{\textbf{3.38 ± 0.15}} & -                    \\ \cline{2-4} 
\multicolumn{1}{c|}{}                                                                                & \multicolumn{1}{c|}{Fake Curve} & \multicolumn{1}{c|}{3.00 ± 0.19}          & -                    \\ \hline
\multicolumn{1}{c|}{\multirow{2}{*}{\textbf{\begin{tabular}[c]{@{}c@{}}Speed\\ Only\end{tabular}}}}  & \multicolumn{1}{c|}{Real Curve} & \multicolumn{1}{c|}{-}                    & \textbf{3.37 ± 0.25} \\ \cline{2-4} 
\multicolumn{1}{c|}{}                                                                                & \multicolumn{1}{c|}{Fake Curve} & \multicolumn{1}{c|}{-}                    & 3.21 ± 0.19          \\ \hline
\multicolumn{1}{c|}{\textbf{Pitch+Speed}}                                                            & \multicolumn{1}{c|}{Real Curve} & \multicolumn{1}{c|}{\textbf{3.18 ± 0.15}} & \textbf{3.41 ± 0.14} \\ \hline
\end{tabular}
\label{sub2_new}
\end{table}

\textbf{Controllability Test.} The second test assesses the controllability of the proposed ControlVC method. We do not include the two baselines in this test due to their poor audio quality in the previous test. 
The participants are presented with uncontrolled and controlled conversion results, along with a figure of the corresponding control curve(s). The participants are then asked to assess how accurately the curve describes the change of pitch or speed between the uncontrolled and controlled conversions on a scale of 1 to 5, with 1 being ``not at all accurate", 3 being``moderately accurate" and 5 being ``very accurate". 
Each participant is asked to complete 6 rounds of tests, with each round containing a pitch control, a speed control and a pitch+speed control of the same source-target reference pair. Participants are allowed to complete more rounds. In total, participants completed 169 rounds of tests, resulting in 676 ratings.

In single-factor control conversions, the presented control curve has a 15\% chance of being a fake curve, which is a flipped or circularly shifted version of the actual one used. This provides us baseline ratings of no controllability for comparison. Table~\ref{sub2_new} shows the assessment results. 
A paired t-test shows that our proposed method achieves a statistically significantly higher MOS rating than the baseline does on pitch control ($p<0.01$). 
On speed control, the MOS difference is more subtle but still statistically significant ($p<0.01$). As some utterances are short, it might be more difficult for the subjects to assess speed controllability.
The differences of MOS ratings between the single-control (pitch-only or speed-only) and pitch+speed control is not statistically significant ($p=0.07$ for pitch; $p=0.43$ for speed).
This suggests that our system is able to control both factors simultaneously without significant quality degradation.

\section{Conclusions}

In this paper, we proposed a controllable voice conversion system named ControlVC, which allows users to impose time-varying controls on pitch and speed in voice conversion. The converted utterance maintains the source utterance's linguistic content, mimics the target speaker's timbre, and sounds natural while following the user input pitch and/or speed controls. 
Both subjective and objective evaluation results suggest that ControlVC is able to perform pitch and speed control while producing high-quality conversions.



\bibliographystyle{IEEEtran}
\bibliography{refs}

\begin{thebibliography}{10}
\providecommand{\url}[1]{#1}
\csname url@samestyle\endcsname
\providecommand{\newblock}{\relax}
\providecommand{\bibinfo}[2]{#2}
\providecommand{\BIBentrySTDinterwordspacing}{\spaceskip=0pt\relax}
\providecommand{\BIBentryALTinterwordstretchfactor}{4}
\providecommand{\BIBentryALTinterwordspacing}{\spaceskip=\fontdimen2\font plus
\BIBentryALTinterwordstretchfactor\fontdimen3\font minus \fontdimen4\font\relax}
\providecommand{\BIBforeignlanguage}[2]{{%
\expandafter\ifx\csname l@#1\endcsname\relax
\typeout{** WARNING: IEEEtran.bst: No hyphenation pattern has been}%
\typeout{** loaded for the language `#1'. Using the pattern for}%
\typeout{** the default language instead.}%
\else
\language=\csname l@#1\endcsname
\fi
#2}}
\providecommand{\BIBdecl}{\relax}
\BIBdecl

\bibitem{Review}
B.~Sisman, J.~Yamagishi, S.~King, and H.~Li, ``An overview of voice conversion and its challenges: From statistical modeling to deep learning,'' \emph{IEEE Trans. ASLP}, vol.~29, pp. 132--157, 2020.

\bibitem{vcc20}
Z.~Yi, W.~Huang, X.~Tian, J.~Yamagishi, R.~K. Das, T.~Kinnunen, Z.~Ling, and T.~Toda, ``{Voice Conversion Challenge 2020}: Intra-lingual semi-parallel and cross-lingual voice conversion,'' \emph{arXiv:2008.12527}, 2020.

\bibitem{emotionbook}
B.~Schuller and A.~Batliner, \emph{Computational paralinguistics: emotion, affect and personality in speech and language processing}.\hskip 1em plus 0.5em minus 0.4em\relax John Wiley \& Sons, 2013.

\bibitem{Carlson1989}
R.~Carlson, A.~Friberg, L.~Fryden, B.~Granstrm, and J.~Sundberg, ``Speech and music performance: Parallels and contrasts,'' \emph{Contemporary Music Review}, vol.~4, pp. 391--404, 1989.

\bibitem{qian2020f0}
K.~Qian, Z.~Jin, M.~Hasegawa-Johnson, and G.~J. Mysore, ``F0-consistent many-to-many non-parallel voice conversion via conditional {Autoencoder},'' in \emph{Proc. {INTERSPEECH}}, {Incheon, Korea}, 2022, pp. 6284--6288.

\bibitem{RN5}
S.~Wang, D.~Kostadinov, and D.~Borth, ``Zero-shot voice conversion via self-supervised prosody representation learning,'' in \emph{Proc. {IJCNN}}, {Padua, Italy}, 2022, pp. 01--08.

\bibitem{li2021starganv2}
Y.~A. Li, A.~Zare, and N.~Mesgarani, ``{StarGANv2-VC}: A diverse, unsupervised, non-parallel framework for natural-sounding voice conversion,'' in \emph{Proc. {INTERSPEECH}}, {Brno, Czechia}, 2021.

\bibitem{chen2010gmm}
L.~Chen, Z.~Ling, W.~Guo, and L.~Dai, ``{GMM-based} voice conversion with explicit modelling on feature transform,'' in \emph{Proc. {ISCSLP}}, {Tainan, Taiwan}, 2010, pp. 364--368.

\bibitem{sun2015voice}
L.~Sun, S.~Kang, K.~Li, and H.~Meng, ``Voice conversion using deep bidirectional long short-term memory based recurrent neural networks,'' in \emph{Proc. ICASSP}, {South Brisbane, Australia}, 2015, pp. 4869--4873.

\bibitem{nercessian2021end}
S.~Nercessian, ``End-to-end zero-shot voice conversion using a {DDSP} vocoder,'' in \emph{Proc. {WASPAA}}, {New Paltz, NY, USA}, 2021, pp. 1--5.

\bibitem{kameoka2018stargan}
H.~Kameoka, T.~Kaneko, K.~Tanaka, and N.~Hojo, ``{StarGAN-VC}: Non-parallel many-to-many voice conversion using star generative adversarial networks,'' in \emph{Proc. {SLT}}, Athens, Greece, 2018, pp. 266--273.

\bibitem{lin2021fragmentvc}
Y.~Y. Lin, C.~Chien, J.~Lin, H.~Lee, and L.~Lee, ``{FragmentVC}: Any-to-any voice conversion by end-to-end extracting and fusing fine-grained voice fragments with attention,'' in \emph{Proc. ICASSP}, Toronto, Canada, 2021, pp. 5939--5943.

\bibitem{casanova2022yourtts}
E.~Casanova, J.~Weber, C.~D. Shulby, A.~C. Junior, E.~G{\"o}lge, and M.~A. Ponti, ``{Y}our{TTS}: Towards zero-shot multi-speaker {TTS} and zero-shot voice conversion for everyone,'' in \emph{Proc. ICML}, Hyderabad, India, 2022, pp. 2709--2720.

\bibitem{nguyen2022nvc}
B.~Nguyen and F.~Cardinaux, ``{NVC-Net}: End-to-end adversarial voice conversion,'' in \emph{Proc. ICASSP}, Singapore, 2022, pp. 7012--7016.

\bibitem{RN14}
K.~Qian, Y.~Zhang, S.~Chang, M.~Hasegawa-Johnson, and D.~Cox, ``Unsupervised speech decomposition via triple information bottleneck,'' in \emph{Proc. ICML}, Virtual, 2020, pp. 7836--7846.

\bibitem{RN8}
A.~Polyak, Y.~Adi, J.~Copet, E.~Kharitonov, K.~Lakhotia, W.-N. Hsu, A.~Mohamed, and E.~Dupoux, ``Speech resynthesis from discrete disentangled self-supervised representations,'' \emph{arXiv:2104.00355}, 2021.

\bibitem{RN18}
L.~Chen and A.~Rudnicky, ``Fine-grained style control in transformer-based text-to-speech synthesis,'' in \emph{Proc. ICASSP}, Singapore, 2022, pp. 7907--7911.

\bibitem{RN11}
F.~Charpentier and M.~Stella, ``Diphone synthesis using an overlap-add technique for speech waveforms concatenation,'' in \emph{Proc. ICASSP}, Tokyo, Japan, 1986, pp. 2015--2018.

\bibitem{hifigan}
J.~Kong, J.~Kim, and J.~Bae, ``{HiFi-GAN}: Generative adversarial networks for efficient and high fidelity speech synthesis,'' \emph{Advances in Neural Information Processing Systems}, vol.~33, pp. 17\,022--17\,033, 2020.

\bibitem{lee2022duration}
S.~Lee, H.~Noh, W.~Nam, and S.~Lee, ``Duration controllable voice conversion via phoneme-based information bottleneck,'' \emph{IEEE Trans. ASLP}, vol.~30, pp. 1173--1183, 2022.

\bibitem{lian2021towards}
Z.~Lian, R.~Zhong, Z.~Wen, B.~Liu, and J.~Tao, ``Towards fine-grained prosody control for voice conversion,'' in \emph{Proc. ISCSLP}, Hong Kong, 2021, pp. 1--5.

\bibitem{choi2021neural}
H.-S. Choi, J.~Lee, W.~Kim, J.~Lee, H.~Heo, and K.~Lee, ``Neural analysis and synthesis: Reconstructing speech from self-supervised representations,'' \emph{Advances in Neural Information Processing Systems}, vol.~34, pp. 16\,251--16\,265, 2021.

\bibitem{kasi2002yet}
K.~Kasi and S.~A. Zahorian, ``Yet another algorithm for pitch tracking,'' in \emph{Proc. ICASSP}, Orlando, FL, USA, 2002, pp. I--361--I--364.

\bibitem{dhariwal2020jukebox}
P.~Dhariwal, H.~Jun, C.~Payne, J.~W. Kim, A.~Radford, and I.~Sutskever, ``Jukebox: A generative model for music,'' \emph{arXiv:2005.00341}, 2020.

\bibitem{RN12}
K.~Qian, Y.~Zhang, S.~Chang, X.~Yang, and M.~Hasegawa-Johnson, ``{AutoVC}: Zero-shot voice style transfer with only {Autoencoder} loss,'' in \emph{Proc. ICML}, Long Beach, USA, 2019, pp. 5210--5219.

\bibitem{RN19}
J.~Yamagishi, C.~Veaux, and K.~MacDonald, ``{CSTR VCTK Corpus}: English multi-speaker corpus for {CSTR Voice Cloning Toolkit} (version 0.92),'' \url{https://doi.org/10.7488/ds/2645}, University of Edinburgh, The Centre for Speech Technology Research ({CSTR}), 2019.

\bibitem{zolzer2002dafx}
U.~Z{\"o}lzer, X.~Amatriain, D.~Arfib, J.~Bonada, G.~De~Poli, P.~Dutilleux, G.~Evangelista, F.~Keiler, A.~Loscos, D.~Rocchesso \emph{et~al.}, \emph{{DAFX}: Digital audio effects}.\hskip 1em plus 0.5em minus 0.4em\relax John Wiley \& Sons, 2002.

\bibitem{fairseq}
{M}eta {Research}, ``{F}acebook {AI} {R}esearch {Sequence-to-Sequence Toolkit} written in {P}ython,'' \url{https://github.com/facebookresearch/fairseq}, GitHub.

\bibitem{Nagrani17}
A.~Nagrani, J.~S. Chung, and A.~Zisserman, ``{VoxCeleb}: a large-scale speaker identification dataset,'' in \emph{Proc. INTERSPEECH}, Stockholm, Sweden, 2017.

\bibitem{RN2}
V.~Panayotov, G.~Chen, D.~Povey, and S.~Khudanpur, ``{LibriSpeech}: an {ASR} corpus based on public domain audio books,'' in \emph{Proc. ICASSP}, South Brisbane, Australia, 2015, pp. 5206--5210.

\bibitem{autovccode}
K.~Qian, ``{PyTorch} implementation of {AutoVC}: Zero-shot voice style transfer with only {Autoencoder} loss,'' \url{https://github.com/auspicious3000/autovc}, GitHub.

\bibitem{RN15}
G.~Saon, H.-K.~J. Kuo, S.~Rennie, and M.~Picheny, ``The {IBM} 2015 english conversational telephone speech recognition system,'' \emph{arXiv:1505.05899}, 2015.

\bibitem{wang2003word}
Y.-Y. Wang, A.~Acero, and C.~Chelba, ``Is word error rate a good indicator for spoken language understanding accuracy,'' in \emph{Proc. ASRU}, Virgin Islands, 2003, pp. 577--582.

\bibitem{Resemblyzer}
{Resemble AI}, ``A {P}ython package to analyze and compare voices with deep learning,'' \url{https://github.com/resemble-ai/Resemblyzer}, GitHub.

\bibitem{streijl2016mean}
R.~C. Streijl, S.~Winkler, and D.~S. Hands, ``Mean opinion score ({MOS}) revisited: methods and applications, limitations and alternatives,'' \emph{Multimedia Systems}, vol.~22, no.~2, pp. 213--227, 2016.

\end{thebibliography}

\end{document}